\documentclass[prl,twocolumn,showpacs,preprintnumbers,amsmath,amssymb]{revtex4}
\usepackage{graphicx}% Include figure files
\usepackage{dcolumn}% Align table columns on decimal point
\usepackage{bm}% bold math

\begin{document}

\newcommand*{\cm}{cm$^{-1}$\,}
\newcommand*{\Tc}{T$_c$\,}

%\reprint{APS/123-QED}

\title{Probing superconducting energy gap from infrared spectroscopy on a Ba$_{0.6}$K$_{0.4}$Fe$_2$As$_2$ single crystal with T$_c$=37 K}% Force line breaks with \\

\author{G. Li}
\author{W. Z. Hu}
\author{J. Dong}
\author{Z. Li}
\author{P. Zheng}
\author{G. F. Chen}
\author{J. L. Luo}
\author{N. L. Wang}

\affiliation{Beijing National Laboratory for Condensed Matter
Physics, Institute of Physics, Chinese Academy of Sciences,
Beijing 100190, China}

%\date{March 26, 2008}% It is always \today, today,

\begin{abstract}

We performed optical spectroscopy measurement on a superconducting
Ba$_{0.6}$K$_{0.4}$Fe$_2$As$_2$ single crystal with T$_c$=37 K.
Formation of the superconducting energy gaps in the far-infared
reflectance spectra below T$_c$ is clearly observed. The gap
amplitudes match well with the two distinct superconducting gaps
observed in angle-resolved photoemission spectroscopy experiments
on different Fermi surfaces. We determined absolute value of the
penetration depth at 10 K as $\lambda\simeq2000 \AA$. A spectral
weight analysis shows that the Ferrell-Glover-Timkham sum rule is
satisfied at low energy scale, less than 6$\Delta$.
\end{abstract}

\pacs{74.70.-b, 74.62.Bf, 74.25.Gz}

% PACS, the Physics and Astronomy
% Classification Scheme.

%\keywords{Suggested keywords}%Use showkeys class option if keyword
                              %display desired
\maketitle

The energy gap created by the pairing of electrons is the most
important parameter of a superconductor. Probing the pairing
energy gap is crucial for elucidating the mechanism of
superconductivity. For conventional superconductors, infrared
spectroscopy is a standard technique to probe the superconducting
energy gap, as the electromagnetic radiation below the gap energy
2$\Delta$ could not be absorbed.\cite{Tinkham1} However, detecting
superconducing energy gap by infrared spectroscopy is not always
straightforward. For example, in the case of high-T$_c$ cuprates
it has been a long standing controversial issue whether the
superconducting gap could be detected from ab-plane infrared
spectra, as it was argued that the ab-plane of the cuprates is in
the clean limit, and as a consequence the pairing gap could not be
seen.\cite{Kamaras}

The recent discovery of superconductivity in FeAs-based
RFeAsO$_{1-x}$F$_x$ (R=rare earth elements like La,Ce,Pr,Nd,Sm and
etc.)\cite{Kamihara08,Chen1,XHChen,Ren1} and (A,K)Fe$_2$As$_2$
(A=Ba, Sr)\cite{Rotter,Chen2,Sasmal} has generated new excitement
in superconductivity community because they represent a new class
of high temperature superconductors. It raises the question
whether the paring mechanism in the new systems is conventional,
or related to that in cuprates. With the success of the growth of
single crystals in the FeAs-based superconductors,\cite{Ni,Chen3}
it is important to investigate the fundamental properties of the
new systems. Such studies are expected to shed new light on the
high temperature superconductivity in cuprates.

In this letter we present an infrared study on a superconducting
Ba$_{0.6}$K$_{0.4}$Fe$_2$As$_2$ single crystal with T$_c$=37 K. We
provide clear evidence that the superconducting gap is present in
the optical reflectance spectra with a s-wave-like pairing
lineshape. From the onset absorption in optical conductivity, the
gap is found to be close to 2$\Delta$=150 \cm, however, from the
peak position in the ratio of the R$_s$(10 K)/R$_n$(45 K) (where
the subscript s stands for superconducting state, n for normal
state) which reflects a more steep drop above this frequency in
optical reflectance relative to the normal state, a different gap
amplitude of 200 \cm is seen. Those two different values match
well with the two distinct superconducting gaps observed in
angle-resolved photoemission spectroscopy (AREPS) experiments on
different Fermi surfaces.\cite{Ding,Zhao} The ability to observe
clearly pairing gaps in infrared spectra indicates that the
material is in the dirty limit. The penetration depth for
T$\ll$T$_c$ is estimated to be $\lambda\simeq2000 \AA$.

High-quality single crystals of Ba$_{0.6}$K$_{0.4}$Fe$_2$As$_2$
were grown by a FeAs flux method.\cite{Chen3} Figure 1 shows the
temperature dependence of the dc resistivity and ac
susceptibility. A sharp superconducting transition is seen at
T$_c$=37 K. The optical measurements were performed on a
combination of Bruker IFS 66v/s and 113v spectrometers on newly
cleaved surfaces. An \textit{in-situ} gold and aluminium
overcoating technique was used for the experiment. Optical
conductivity was derived from Kramers-Kronig transformation of the
reflectance.

\begin{figure}[b]
\includegraphics[width=7.8cm,clip]{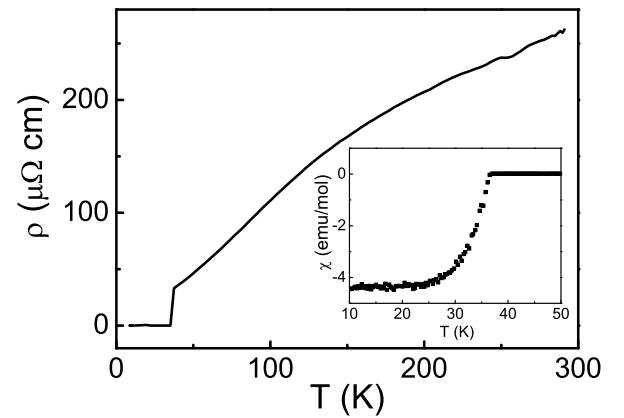}
\caption{The dc resistivity as a function of temperature for a
Ba$_{0.6}$K$_{0.4}$Fe$_2$As$_2$ single crystal. Inset shows the ac
susceptibility of the sample. Sharp superconducting transition
occurs at T$_c$=37 K.}
\end{figure}

Figure 2 shows the reflectance spectra R($\omega$) for the crystal
at different temperatures. The inset shows the spectra over broad
energy range up to 25000 \cm, while the main panel is the expanded
plot below 800 \cm. R($\omega$) exhibits a metallic response in
both frequency and temperature. At 10, 27 and 45 K, the
reflectance curves almost overlap with each other above 300 \cm.
However, a sudden upturn R($\omega$) develops below T$_c$ at low
frequencies. This is a strong indication for the formation of an
superconducting energy gap due to the pairing of electrons.

\begin{figure}[t]
\includegraphics[width=7.8cm,clip]{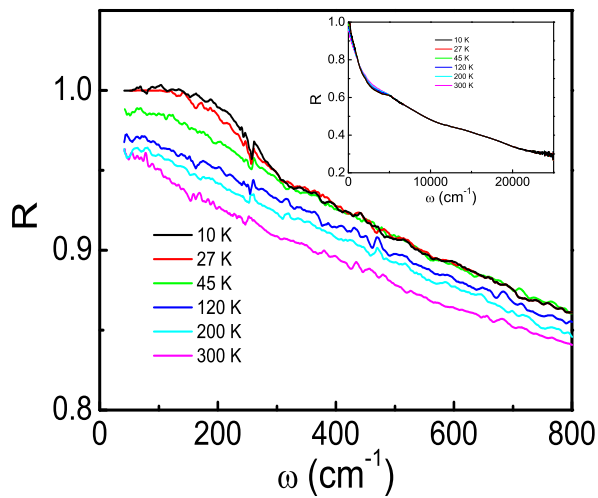}
\caption{(Color online) T-dependent $R(\omega$) curves in the
far-infrared region. The inset shows $R(\omega$) over broad
frequencies up to 25000 \cm.}
\end{figure}

It is well known that for a s-wave BCS superconductor with an
isotropic superconducting energy gap, the reflectivity approaches
unity below 2$\Delta$.\cite{Ortolani,Degiorgi} The R($\omega$)
curve at 10 K is almost flat below 150 \cm (18 meV), thus being
very similar to the lineshape for a s-wave superconductor.
However, above this frequency R($\omega$) develops a downward
curvature, and its magnitude becomes slightly lower than unity,
suggesting that week absorption already exists. The decrease of
R($\omega$) becomes steep above 200 \cm.

Figure 3 shows the optical conductivity $\sigma_1(\omega$) derived
from the Kramers-Kronig transformation of reflectance spectra. The
conductivity values at very low frequencies continue to increase
with decreasing temperature in the normal state, being consistent
with dc resistivity measurement. Below T$_c$, the curves at 27 K
and 10 K decreases steeply near 300 \cm. The conductivity is
almost zero below roughly 150 \cm, yielding optical evidence for a
s-wave like superconducting energy gap. The onset of the
absorption marks the superconducting energy 2$\Delta\simeq$150
\cm. In recent AREPS experiments on the same batch of single
crystals, two distinct superconducting gaps were observed: a large
gap ($\Delta\simeq$12 meV) on the two small hole-like (centered at
$\Gamma$) and electron-like (centered at M) Fermi surface (FS)
sheets, and a small gap ($\Delta\simeq$6-8 meV) on the large
hole-like FS (centered at $\Gamma$). Both gaps, closing
simultaneously at the bulk T$_c$, are nodeless and nearly
isotropic around their respective FS sheets.\cite{Ding,Zhao} In
comparison with those work, the conductivity onset should
correspond to the small gap observed in ARPES, since optical
absorption should exist when the radiation energy is higher than
this small superconducting pairing gap. Within experimental
uncertainties, the measurement results from the two different
techniques match quite well.

\begin{figure}[t]
\includegraphics[width=8.5cm,clip]{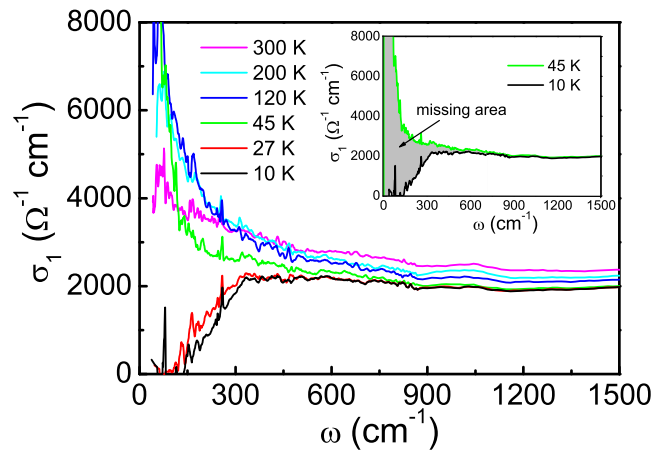}
\caption{(Color online) T-dependent $\sigma_1(\omega$) curves. The
inset shows $\sigma_1(\omega$) at 10 and 45 K. Shaded area
represents the missing area due to the opening of superconducting
energy gap.}
\end{figure}

Figure 4 shows the ratio of the R$_s$(10 K)/R$_n$(45 K). The total
variation exceeds 2$\%$. A peak can be clearly seen at 200 \cm (25
meV). Within BCS framework in the dirty limit, the peak frequency
roughly corresponds to the superconducting energy gap
2$\Delta$.\cite{Degiorgi,Lupi} We noticed that this gap value is
different from the absorption onset in the optical conductivity,
which gives smaller value of 150 \cm. In fact, the frequency at
200 \cm below T$_c$ represents a more steep drop beyond this
energy in optical reflectance relative to the normal state. It is
most likely this peak frequency is related to the large
superconducting gap observed on the two small hole-like and
electron-like FS in ARPES experiment.\cite{Ding} Note that, the
2$\Delta\simeq$200 \cm (25 meV) seen in optics also matches
excellently with the gap $\Delta\simeq$12 meV measured relative to
Fermi level in ARPES.

\begin{figure}[b]
\includegraphics[width=7.8cm,clip]{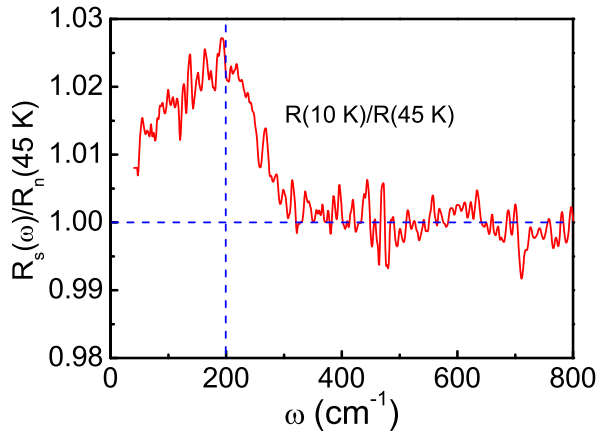}
\caption{(Color online) The reflectance $R(\omega$) at 10 K
normalized to the values at 45 K in the normal state. A peak near
200 \cm is seen.}
\end{figure}

Well below T$_c$, there is a substantial suppresssion in the
low-frequency conductivity due to the formation of superconducting
energy. According to the Ferrell-Glover-Tinkham (FGT) sum
rule,\cite{Ferrell,Tinkham2} the difference between the
conductivity at T$\simeq$$T_c$ and T$\ll$T$_c$ (the so-called
missing area, see the inset of Fig. 1) is related to the formation
of a superconducting condensate,
\begin{equation}
\omega_{ps}^2=8\int_{0^+}^{\omega_c}[\sigma_1(\omega, T\simeq
T_c)-\sigma_1(\omega, T\ll T_c)]. \label{chik}
\end{equation}
where $\omega_{ps}^2=4\pi n_se^2/m^*$ is the square of the
superconducting plasma frequency, $n_s$ is the condensed carrier
density, and $\omega_c$ is the high-frequency cut-off frequency
which should be chosen such that the $\omega_{ps}^2$ converges
smoothly. The penetration depth is related to the superconducting
plasma frequency by $\lambda=c/\omega_{ps}$. Equation (1) states
that the spectral weight lost in $\sigma_1(\omega)$ in the
superconducting state has been transferred to the zero frequency
delta function response of the superconducting condensate. A
direct estimation from the missing area gives $\lambda$=2080
$\AA$.

The superconducting penetration depth can also be estimated from
the imaginary part of the complex conductivity in the
low-frequency limit via\cite{Basov}
\begin{equation}
\lambda(\omega)=c/\omega_{ps}=c/\sqrt{4\pi\omega\sigma_2(\omega)}.
\end{equation}
The determination of $\lambda$ from this equation at low frequency
limit relies only on the imaginary part of conductivity at
T$\ll$T$_c$. When the FGT sum rule is fulfilled, the $\lambda$
value determined from the missing area in the real part of
conductivity should equal to the value obtained by Eq. (2) in the
low-frequency limit.\cite{Homes1} Figure 5 shows the
$\lambda(\omega)$ obtained from above formula, at low frequency
limit, $\lambda\simeq$1950 $\AA$. We find that the value of
$\lambda$ obtained directly from Eq. (1) is close to that obtained
from the imaginary part of conductivity through Eq. (2). The good
agreement (within an accuracy of 5-8$\%$) between the values
obtained from the two different approaches suggests that the FGT
sum rule is satisfied.

\begin{figure}
\includegraphics[width=7.8cm,clip]{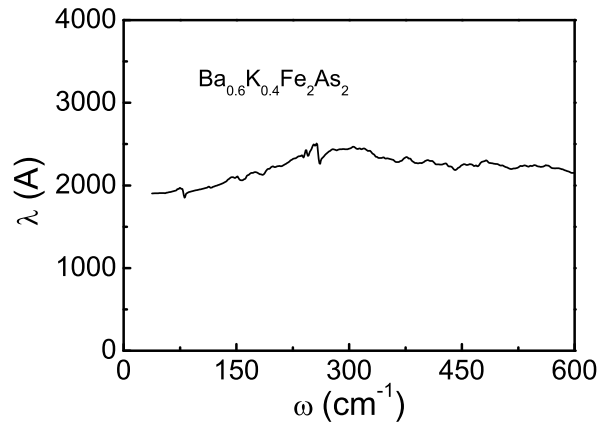}
\caption{Frequency dependent London penetration depth
$\lambda(\omega)=c/\omega_{ps}=c/\sqrt{4\pi\omega\sigma_2(\omega)}$
for Ba$_{0.6}$K$_{0.4}$Fe$_2$As$_2$. Data are at 10 K.}
\end{figure}

It would be interesting to compare the optical response of
FeAs-based supercondutor with those found for the high-T$_c$
cuprate superconductors. Several eminent differences exist: first,
the observation of clear superconducting gap in the far-infrared
spectra suggests that the in-plane superconductivity is in the
dirty limit, i.e. the carrier scattering rate 1/$\tau\geq2\Delta$.
In this case, the spectral weight of itinerant carriers in optical
conductivity distributes in a broad frequency region, however, a
large part of the condensate has been captured by the energy of
2$\Delta$, as can be seen in the inset of Fig. 1. While in the
clean limit case, nearly all spectral weight associated with the
condensate lies below 2$\Delta$, so that no discernable change
appears at 2$\Delta$ across the superconducting transition. It is
still a controversial issue whether the high-T$_c$ cuprates is in
the dirty or clean limit and whether the superconducting gap is
visible in infrared reflectance
spectra.\cite{Kamaras,Homes2,Homes3} It is expected that studies
on the new FeAs-based superconductors may shed light on the
pairing gap feature in the infrared spectra in cuprates.

Secondly, our analysis based on a comparison of the penetration
depth values determined from the missing area in the real part of
conductivity and imaginary part of conductivity at low frequency
limit indicates that the FGT sum rule is satisfied. An inspection
of the inset of Fig. 3 reveals that the missing area extends to
the frequency roughly below 600 \cm, about 3 times larger than the
higher superconducting energy gap 2$\Delta$. This indicates that
the superconducting condensate forms rapidly or the FGT sum rule
is rapidly recovered. This is very different from underdoped
high-T$_c$ cuprates where recovery of the FGT sum rule goes to
very high energy, or the FGT sum rule is even
violated.\cite{Homes2}

Thirdly, the determination of the penetration depth or
equivalently the condensed carrier density enables us to check
whether the well-known scaling behaviors between the condensed
carrier density and T$_c$ still work for the present system. One
of such scaling behaviors is called Uemura
relation,\cite{Uemura1,Uemura1} which states that the superfluid
density scales linearly with the transition temperature,
$\rho_s=\omega_{ps}^2=c^2/\lambda^2\propto T_c$. Uemura relation
works well for the hole-doped cuprates in the underdoped region.
The relatively low value of T$_c$ and low penetration depth in the
present system may fall off the Uemura plot. Homes proposed
another scaling relation, $\rho_s\simeq65\sigma_{dc}T_c$ for BCS
weak coupling case, where $\sigma_{dc}$ is the value just above
T$_c$.\cite{Homes2} The present system seems to fit better to
Homes's scaling relation.

To summarize, we performed infrared spectroscopy measurement on a
superconducting Ba$_{0.6}$K$_{0.4}$Fe$_2$As$_2$ single crystal
with T$_c$=37 K. We observe clearly that the superconducting gap
is present in the optical reflectance spectra with a s-wave
pairing lineshape. The onset absorption in optical conductivity
appears close to 150 \cm, however, a more steep reflectance
decrease relative to the normal state appears at frequency near
200 \cm, leading to a peak position in the ratio of the R$_s$(10
K)/R$_n$(45 K). Those two values match well with the two distinct
superconducting gaps observed in AREPS experiments on different
Fermi surfaces. The ability to observe clearly pairing gaps in
infrared spectra indicates that the material is in the dirty
limit. The penetration depth for T$\ll$T$_c$ is estimated to be
$\lambda\simeq2000 \AA$.

\begin{acknowledgments}
This work is supported by the National Science Foundation of
China, the Knowledge Innovation Project of the Chinese Academy of
Sciences, and the 973 project of the Ministry of Science and
Technology of China.

\end{acknowledgments}

%\bibliography{MgIrB}% Produces the bibliography vi BibTeX.

\end{document}